\def\year{2018}\relax
\documentclass[sigconf, review=false]{acmart}
\usepackage{times}  
\usepackage{helvet}  
\usepackage{courier}  
\usepackage{url}  
\usepackage{graphicx}  
\usepackage{multirow}
\usepackage{booktabs, tabularx}
\usepackage{nicefrac}
\usepackage{balance}
\usepackage{enumitem}

\newcommand\Tt{\rule{0pt}{3.2ex}} 
\newcommand\Tf{\rule{0pt}{2.8ex}}

\begin{document}

\copyrightyear{2019} 
\acmYear{2019} 
\setcopyright{acmlicensed}
\acmConference[UMAP '19]{27th Conference on User Modeling, Adaptation and Personalization}{June 9--12, 2019}{Larnaca, Cyprus}
\acmBooktitle{27th Conference on User Modeling, Adaptation and Personalization (UMAP '19), June 9--12, 2019, Larnaca, Cyprus}
\acmPrice{15.00}
\acmDOI{10.1145/3320435.3320451}
\acmISBN{978-1-4503-6021-0/19/06}

\title{Effect of Values and Technology Use on Exercise:\\ Implications for Personalized Behavior Change Interventions}
\renewcommand{\shorttitle}{Effect of Values and Technology Use on Exercise}

\author{Yelena Mejova and Kyriaki Kalimeri}
\affiliation{%
  \institution{ISI Foundation, Turin, Italy}
}
\email{yelenamejova@acm.org,kalimeri@ieee.org}

\renewcommand{\shortauthors}{Mejova \& Kalimeri}

\begin{abstract}

Technology has recently been recruited in the war against the ongoing obesity crisis; however, the adoption of Health \& Fitness applications for regular exercise is a struggle. In this study, we present a unique demographically representative dataset of 15k US residents that combines technology use logs with surveys on moral views, human values, and emotional contagion. Combining these data, we provide a holistic view of individuals to model their physical exercise behavior. First, we show which values determine the adoption of  Health \& Fitness mobile applications, finding that users who prioritize the value of \emph{purity} and de-emphasize values of \emph{conformity}, \emph{hedonism}, and \emph{security} are more likely to use such apps. Further, we achieve a weighted AUROC of .673 in predicting whether individual exercises, and we also show that the application usage data allows for substantially better classification performance (.608) compared to using basic demographics (.513) or internet browsing data (.546). We also find a strong link of exercise to respondent socioeconomic status, as well as the value of \emph{happiness}. Using these insights, we propose actionable design guidelines for persuasive technologies targeting health behavior modification.

\end{abstract}


\maketitle

\section{Introduction} 

Over the last half-century, the daily occupation-related energy expenditure of US workers has decreased by more than 100 calories, with the proportion of jobs requiring at least moderate intensity physical activity declining from 48\% to 20\% in 2008 \cite{church2011trends}. Sedentary lifestyle -- one that includes TV watching and gaming and lacks vigorous exercise -- 
has also been found to be strongly related to the childhood obesity epidemic \cite{andersen1998relationship}. 

Technology has been an essential factor in changing lifestyles. Over the years, TV has been seen as a replacement of physical activity, a channel for advertisement of nutrient-poor food, and increasing prevalence of ``mindless'' eating \cite{boulos2012obesitv}. The arrival of mobile technology has been shown to contribute to the sedentary behaviors \cite{barkley2016mobile}. However, this has not stopped both entrepreneurs and public health officials from attempting to use new technology to encourage behavior change. Unfortunately, within the industry, user retention is an ongoing struggle, with 62\% of mHealth app publishers report digital solutions with less than 1,000 monthly active users \cite{Research2Guidance}. Indeed, the latest research shows a complex relationship between psychology, technology use, and exercise. Mediation analyses find that increased physical activity associated with health app use is related to feelings of self-efficacy \cite{litman2015mobile}, with yet other insights linking exercise to being extroverted, neurotic, and less agreeable \cite{hausenblas2004relationship,brunes2013personality}, as well as having implications for mental health \cite{rhodes2006personality}. For a better understanding of the relationship between psychology, technology, and exercise, it is necessary to model users of new technologies \cite{cena2018towards} and to design effective health interventions.


This study is a unique view of the interaction between technology use, demographics, and value systems of a representative US population sample, allowing for rich user modeling in the aims of promoting exercise. Just over 15k participants filled in the questionnaires including the following psychometric measures: Moral Foundations \cite{Haidt2004, Haidt2007}, Schwartz Basic Human Values \cite{Schwartz2012}, and Emotional Contagion \cite{doherty1997emotional}. Along with these, 5,008 respondents agreed to allow the capture of their desktop browsing data, whereas another 2,625 allowed to capture their mobile app usage. Using this data, we contribute the following insights on psychological markers of health app use and the actual exercise behavior:

\begin{itemize}[topsep=5pt,itemsep=5pt]
\setlength\itemsep{0em}

\item \textbf{Modeling ``Health \& Fitness'' application use} in relation to psychometric and demographic variables, we find a marked difference in application usage between the two genders, as well as significant negative  relationship between the values of \emph{tradition}, \emph{conformity}, \emph{hedonism}, and \emph{security}, while positive for \emph{purity}.

\item \textbf{Predicting the engagement in physical exercise} via the above variables, as well as browsing and application use data, we show a marked increase in classification performance from baseline demographic model with the addition of psychometric features, as well as application usage data, but with a smaller contribution of desktop browsing data.

\item \textbf{Revealing determinants of exercise} among the types of variable, confirming a significant effect of education and wealth on healthy behaviors \cite{rimal2002association}, as well as showing significant relationships with the view that \emph{health is a choice}, positive association with \emph{happiness} emotional contagion and \emph{stimulation} value, with downloading a Health \& Fitness app being another strong predictor.

\item \textbf{Comparing the exercise behaviors across mobile applications}, we show those tracking a particular kind of exercise, such as running or cycling, are associated with more users reporting exercising regularly, than those for general health tracking or women's health tracking.
\end{itemize}

We conclude with concrete suggestions of employing this knowledge in the design, personalization, and deployment of technologies for an effective lifestyle change and health outcomes intervention.


\section{Related work} 

Technology is now commonly used to monitor behavior and physical activity \cite{ghanvatkar2019user,Kalimeri2010}. 
Digital data from smartphones were initially used as simple activity monitoring sensors \cite{Acampora2013}; however, over the last years, their integration with user-generated content led to more sophisticated personalized interventions aiming at motivating the users to increase their physical activity level and encouraging a healthier lifestyle \cite{opdenAkker2014,Harrington2018}.
Researchers have tried apps with different messaging strategies \cite{yom2017encouraging}, personalized exercise recommendation \cite{tseng2015interactive}, as well as utilizing machine learning via supervised learning \cite{marsaux2016changes,hales2016mixed} and reinforcement learning \cite{yom2017encouraging,rabbi2015mybehavior}. 
Others help users find exercise partners \cite{hales2016mixed}, provide educational materials \cite{alley2016web,short2017different}, and emotional support \cite{vandelanotte2015tayloractive,yom2017encouraging} (see \cite{ghanvatkar2019user} for a recent survey on personalized health interventions). 

To understand the impact of such interventions, researchers examined the role of individual characteristics, attitudes, and lifestyle of users \cite{pratt2012implications,carroll2017uses}, demographic attributes such as gender, age, socioeconomic factors, and technology literacy \cite{baker2000measuring,ernsting2017using,carroll2017uses}.
For instance, using self-reported technology use, \cite{ernsting2017using,carroll2017uses} found that two-thirds of their participants were using a smartphone. This subset was younger, more likely to have a university degree with higher socioeconomic status, and was more likely to engage physical activity. 
A substantial proportion of their population was not engaged in Health \& Fitness apps; however, those who were were more motivated to change or maintain a healthy lifestyle.
A further association was confirmed between smartphone use and health literacy \cite{bailey2015literacy}, and an association with age, with seniors (65 years and older) using digital health at much lower (but steadily increasing) rates \cite{levine2016trends,smith2015us}.
Unlike these previous studies relying on self-reported data based on surveys, our data present a snapshot of desktop and mobile use which provides valuable ground truth for tech-related behaviors (as well as a complementary rich demographic baseline).

Despite being an active research direction, the consideration of the psychological aspects of the individual such as personal views, values, and emotional states, in tech-driven intervention remains hugely unexplored \cite{kerner2017motivational,mclennan2015quality}.
Human values are known to influence people's actions \cite{allport1960study} but received little attention in studying their relationship with attitudes regarding healthy lifestyles. 
Regarding human and moral values in sports, Lee et al. \cite{lee2008relationships} examined the value-expressive function of attitudes and achievement goal theory in predicting the moral attitudes of young athletes. 
Ball et al. \cite{ball2010healthy} studied the individual preferences for social support according to the values system in following a healthy lifestyle.
Apart from these studies, human values were only considered in cases where an individual deviated from a normative of healthy lifestyle \cite{lathia2013smartphones} such as depression \cite{saeb2015mobile} or mental disorders \cite{SANCHEZVALDES20159574}. 
Among works closest to our intention, Lathia et al. \cite{Lathia2017} assessed the relationship between physical activity and happiness via a smartphone app concluding that people that exercise more are happier.
In this study, we contribute a unique combination of psychometric measures, spanning morals, values, and emotional contagion, to better understand technology use and engagement in physical activity.

\section{Data Collection} 

Data presented in this study spans 15,021 subjects in the United States of America, selected using probabilistic, representative sampling methodology, all of whom were incentivised to participate. 
After receiving informed consent from all participants for the collection, storage, and analysis of the data, as well as the acceptance of the privacy policy\footnote{https://www.researchnow.com/privacy-policy/}, we administered a series of questionnaires to gather demographic and psychometric data.
Also, we asked the participants for the access to either their basic mobile or desktop data for one month, resulting in desktop activity data for 5,008 people (2,823 women) and mobile activity data for 2,625 people (1,544 women).
The latter subset with activity data has been discussed in \cite{kalimeri2018predicting}.
Below we describe the data collected and used in this study\footnote{For privacy considerations, the data will be made available upon request, exclusively for the scientific community.}.

\subsection{Demographics}

The intake survey covered basic demographic factors (age,  gender,  ethnicity), 
geographic factors (home location,  expressed at the zip code level),  socioeconomic factors (educational level,  marital status,  parenthood,  wealth,  income),   health-related factors (exercise,  smoke, and weight issues) and political orientation. 
Table \ref{tab:Demog} presents the complete list of the demographic information gathered, along with the respective range of values for all the 15,021 participants.

\begin{table*}
\small
\centering
\caption{Complete list of the demographic attributes collected and their respective ranges for the entire sample of 15,021 participants.}
\label{tab:Demog}
\renewcommand{\arraystretch}{1.2}
\begin{tabularx}{\textwidth}{p{0.12\linewidth}p{0.17\linewidth}p{0.16\linewidth}p{0.12\linewidth}p{0.17\linewidth}p{0.10\linewidth}}\noalign{\smallskip}\noalign{\smallskip}
\toprule
\textbf{Attribute} &{\textbf{Demographic Variables}}&  \textbf{Sample size}&\textbf{Attribute} &{\textbf{Demographic Variables}}&  \textbf{Sample size}\\
& \textbf{Range} &{($N=15,021$)}& & \textbf{Range} &{($N=15,021$)}\\
\midrule
\textbf{Age} &18-24 & 1,636 (10.8\%) & 
\textbf{Political Party}&Democrat & 6,227 (41.4\%) \\
&25-34 & 2,583 (17.1\%) & & Republican & 4,455 (29.6\%) \\
&35-49 & 3,770 (25\%) & &Libertarian & 429 (2.8\%) \\
&50-54 & 1,642 (10.9\%) & &Independent & 3,910 (26\%) \\
&55-64 & 2,707 (18\%) &&&\\
& 65+ & 2,683 (17.8\%) &&&\\

\textbf{Education}\Tt&College Graduate & 4,854 (32.3\%) & 
\textbf{Wealth}&50k or less & 5,520 (36.7\%) \\
&Post Graduate & 3,409 (22.6\%) & & 50k-100k & 2,087 (13.8\%) \\
&Some College & 3,810 (25.3\%) & &100k-250k & 2,375 (15.8\%) \\
&High-school & 1,832 (12.1\%) & &250k-500k & 2,166 (14.4\%) \\
&Trade School & 949 (6.3\%) & &500k-1000k&  1,627 (10.8\%) \\
&&& & 1000k or more & 1,246 (8.2\%) \\

\textbf{Ethnicity}\Tt &Asian  & 669 (4.4\%) & 
\textbf{Weight Issues} & No & 8,709 (57.9\%) \\
& African American & 1,761 (11.7\%) & & Yes & 6,312 (42\%) \\
&White & 11,042 (73.5\%) \\
& Hispanic & 1,335 (8.8\%) \\

\textbf{Exercise} \Tt & No & 6,631 (44.1\%) & 
\textbf{Parent} & No & 5,613 (37.4\%) \\
& Yes & 8,390 (55.8\%) & &Yes & 9,408 (62.6\%) \\

\textbf{Gender} \Tt& Female & 8,409 (55.9\%) & 
\textbf{Smoker} & No & 13,150 (87.5\%) \\
& Male & 6,612 (44.1\%) & & Yes & 1,871 (12.4\%) \\

\textbf{Income} \Tt &20k or less & 1,384 (9.2\%) & 
\textbf{Marital Status} &Divorced & 1,409 (9.3\%) \\
&20k-30k & 1,389 (9.2\%) & & Single & 3,509 (23.3\%) \\
&30k-50k & 2,785 (18.5\%) & &Married & 8,037 (53.5\%) \\
&50k-75k & 3,246 (21.6\%) & &Living Together & 1,444 (9.6\%) \\
&75k-100k & 2,601 (17.3\%) &&&\\
&100k-150k & 2,386 (15.8\%) & 
\textbf{High Blood}& No&12,025 (80\%)\\
&150k-200k & 745 (4.9\%) & \textbf{Pressure} &Yes&2,996 (20\%)\\
& 200k or more & 485 (3.2\%) &&&\\

\bottomrule
\end{tabularx}
\end{table*}

\subsection{Psychometric Measures}

\paragraph*{Moral Foundations}

To measure the values of the participants, we employ the Moral Foundation Theory \cite{Haidt2004, Haidt2007} which we operationalized via the Moral Foundations Questionnaire (MFQ) \cite{Graham2011}, a validated measure of the degree to which individuals endorse each of five dimensions:

\begin{itemize}[topsep=3pt,itemsep=0pt]
\item\emph{care/harm},  basic concerns for the suffering of others, including virtues of caring and compassion;
\item\emph{fairness/cheating},  concerns about unfair treatment,  inequality,  and more abstract notions of justice;
\item\emph{loyalty/betrayal},  concerns related to obligations of group membership,  such as loyalty,  self-sacrifice, and vigilance against betrayal;
\item \emph{authority/subversion},  concerns related to social order and the obligations of hierarchical relationships
like obedience,  respect,  and proper role fulfillment;
\item \emph{purity/degradation},  concerns about physical and spiritual contagion,  including virtues of chastity,  wholesomeness, and control of desires.
\end{itemize}

The questionnaire is based on self-assessment evaluations and consists of 30 items, resulting in a unique numerical value from 0-30 per person. According to the MFQ, six items (on a 6-point Likert scale) per foundation were averaged to produce the individuals' scores on each of the five foundations.

\paragraph*{Schwartz Basic Human Values}
We assess the \textit{Schwartz human values} employing the Portrait Values Questionnaire \cite{Schwartz2012}, whose validity across cultures is validated in studies performed on 82 countries and samples belonging to highly diverse geographic,  cultural,  linguistic,  religious,  age,  gender,  and occupational groups. The questionnaire is based on self-assessments resulting in a numerical value per person for each of the ten basic values:

\begin{itemize}[topsep=3pt,itemsep=0pt]
\item \emph{self-direction},  independent thought,  action-choosing,  creating,  exploring;
\item \emph{stimulation},  need for variety and stimulation to maintain an optimal level of activation;
\item \emph{hedonism},   related to organismic needs and the pleasure associated with satisfying them;
\item \emph{achievement},  personal success through demonstrating competence according to social standards;
\item \emph{power},  the attainment or preservation of a dominant position within the more general social system;
\item \emph{security},   safety,  harmony,  and stability of society,  of relationships,  and self;
\item \emph{conformity},  restraint of actions,  inclinations,  and impulses likely to upset or harm others and violate social expectations or norms;
\item \emph{tradition},  symbols and practices or groups that represent their shared experience and fate;
\item \emph{benevolence},   concern for the welfare of close others in everyday interaction;
\item \emph{universalism},  this value type includes the former maturity value type, including understanding,  appreciation,  tolerance,  and protection for the welfare of all people and nature.
\end{itemize}

The questionnaire is based on self-assessment evaluations on a 7-point Likert scale.
Following \cite{Schwartz2012}, we average the respective items per value, and we account for individual differences.
The above ten values can be clustered into four higher order values,  so-called quadrant values and into two dimensions, as the sum of the individual items of which they consist:
\emph{Openness to change} (self-direction,  stimulation) vs. \emph{Conservation} (security,  conformity,  tradition) and \emph{Self-enhancement} (universalism,  benevolence)  vs. \emph{Self-transcendence} (power,  achievement).
Therefore,  the first dimension captures the conflict between values that emphasize the independence of thought,  action,  and feelings and readiness for change and the values that highlight order,  self-restriction,  preservation of the past,  and resistance to change. The second dimension captures the conflict between values that stress concern for the welfare and interests of others and values that emphasize the pursuit of one's interests and relative success and dominance over others. Hedonism shares elements of both openness to change and self-enhancement.

\paragraph*{Emotional Contagion}

Emotional contagion is the phenomenon that individuals tend to feel emotions, such as happiness, or sadness, triggered by the feelings expressed by the people with whom they interact \cite{hatfield1993emotional}. In this study, we employ the well-established emotional contagion scale (EC) \cite{doherty1997emotional}. The 15-item questionnaire is based on self-assessment evaluations on a 5-point Likert scale.
It assesses mimetic tendency to five basic emotions (love, happiness, fear, anger, and sadness), measuring the individual differences in susceptibility to ``catching'' and empathizing the emotions of others.

\subsection{Digital Data}
\paragraph*{Desktop Browsing Data}

For the participants who permitted the logging of their desktops' web browsing data, 5,008 in total, we capture: (i) the domain names, and (ii) the average time spent online and (iii) the number of visits per day on each domain. All this information is aggregated by day, and only the domain names (and not the page or section of the websites) are stored, to ensure the privacy of the participants. Users with fewer than $N = 30$ unique domains are discarded. We then assign to each domain name a category label according to its content \cite{Domain_cat2015}.

\paragraph*{Mobile Data}
\label{sec:mob_data}

Participants are also asked to download an application which, upon agreement with the privacy policy, logs their web browsing activity and application usage, and 2,625 agreed to be tracked. 
\begin{itemize}
\item \emph{Application Data.} Application usage was captured whenever the application was running in the ``foreground''. Foreground usage means an application is open on someone's device,  regardless of whether the application is currently being engaged with or not.
Application usage data for each participant included records of the date and time stamp, the local time zone, and time spent on the application (in seconds). Moreover, we assigned to each application the category label provided by the Google Play Store\footnote{The assignment was performed parsing the application data from the Google Play Store using the following \href{https://github.com/danieliu/play-scraper}{project}.}. 
\item \emph{Mobile browsing Data.}  URL data was captured from the native browser on the subject's device (not any 3rd party browsers). URLs for both secure and non-secure traffic were captured,  though only the URL domain was stored in consideration of privacy.  Similar to the desktop browsing data, users with a number of visits fewer than $N = 30$ unique domains are discarded from the analysis leaving us with a total of 2,406 participants.
The domains are classified as above for the Desktop users \cite{Domain_cat2015}. 
\end{itemize}
 
Noteworthy is the fact that ``Mobile'' and ``Desktop'' browsing data provide the same information, they only express different modes of web navigation, i.e. mobile vs desktop. See \cite{kalimeri2018predicting} for a detailed description of the data.

\section{Health App Use} 
\label{sec:healthappuse}

We begin by examining the usage of mobile applications (apps) in the Health \& Fitness category. These apps include those associated with particular wearables like \emph{Fitbit} and \emph{Garmin Connect}, activity trackers like \emph{MapMyRun}, \emph{RunKeeper}, \emph{Nike+ Run Club}, and weight management including \emph{Lose It!} and \emph{WW (Weight Watchers)}. The only demographic variable associated with getting such an app (more precisely, opening it at least once in the time of observation), is gender, with females 45\% more likely to get one than males. Though the gender division is not evenly distributed across applications, with those marketed for tracking running activity (such as \emph{RunKeeper} and \emph{MapMyRun}) being 24.9\% more likely to be adopted by females, whereas those for walking (\emph{Walkroid} and \emph{MapMyWalk}) are 101\% more likely (that is, twice as likely). The distinction is even greater for weight loss applications, with females 168\% more likely to adopt one than males. Notably, these gender distinctions have not been revealed in recent surveys \cite{krebs2015health,ernsting2017using}.

\begin{figure}[t]
\centering
\includegraphics[width=0.85\linewidth]{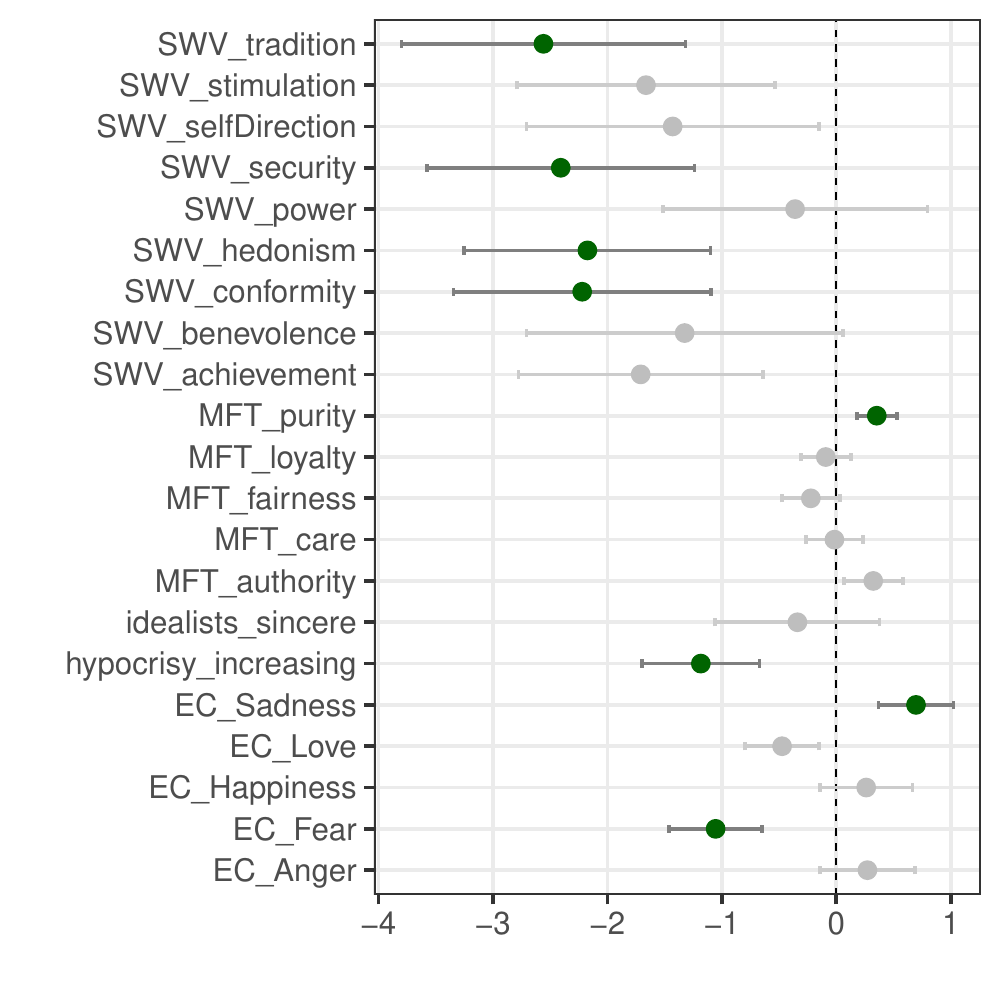}
\caption{Coefficients of linear model predicting Health \& Fitness app use, with those significant at $p < 0.05$ bolded in green.}
\label{fig:coeff_app_use}
\end{figure}

Considering the psychometric attributes, we run a linear model ($n=2620$) to predict the adoption of any Health \& Fitness application, with the coefficients plotted in Figure \ref{fig:coeff_app_use}, whiskers marking 95\% confidence intervals, and those significant at $p < 0.05$ bolded in green. We find a negative relationship with values associated with \emph{tradition}, \emph{security}, \emph{hedonism}, and \emph{conformity}, as well as with concerns about \emph{hypocrisy} increasing in the society, and a positive relationship with \emph{purity} value. Emotional contagion results show a positive relationship with \emph{sadness} but a negative with \emph{fear}. These trends point to people less concerned about societal traditions, who are less influenced by caution or fear, and those striving towards physical or spiritual purity.

\begin{figure}[t]
\centering
\includegraphics[width=0.50\linewidth]{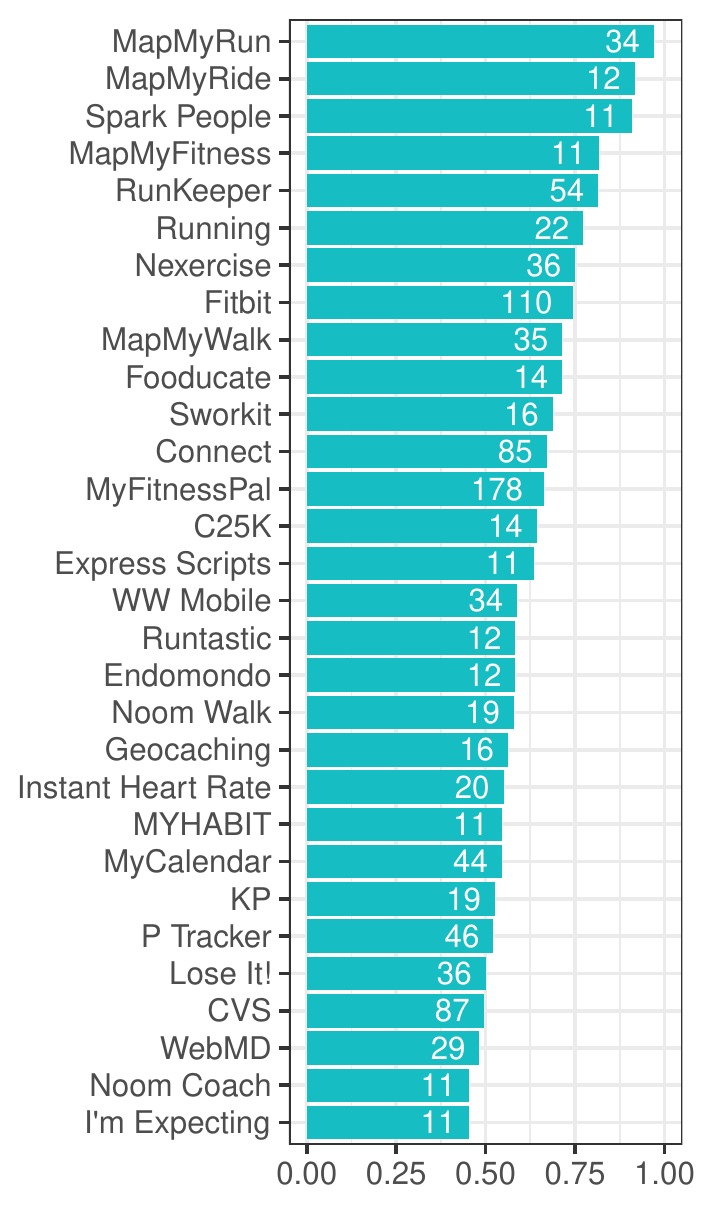}
\caption{Applications in the Health \& Fitness category ordered by the proportion of respondents reporting exercising regularly, with the $n$ shown in white.}
\label{fig:apps_for_exercise}
\end{figure}

Finally, we ask which health applications are most associated with self-reported exercise (defined in more detailed in the following section). To answer this question, we consider all applications in the Health \& Fitness category having at least ten users in our dataset and compute the proportion of such users who self-report exercising. The top 30 apps are shown in Figure \ref{fig:apps_for_exercise}, along with the number of users the proportion is based on.

Running, cycling, and walking tracking applications dominate the top, as well as \emph{Spark People}, which provides a combination of weight loss and fitness (although the proportion is based on eleven respondents). Towards the bottom, we find generic health resources like \emph{WebMD}, as well as pregnancy apps (\emph{I'm Expecting}). Thus, we observe applications with an explicit activity to perform are better at supporting regular exercise than, say, more generic pedometers or health trackers.

\section{Modeling Exercise} 

Thus, we find value determinants in our study subjects' willingness to use the health applications, but we are interested in whether such knowledge would help understand the step to exercise. Indeed, we find that respondents who have downloaded such applications are 42\% more likely to say they exercise ($p < 0.001$). However, as we illustrate in the following sections, exercise behavior has a multifaceted nature beyond health app usage.

\subsection{Demographics of Exercise}

In this study, we operationalize health-related activities of the user via the questionnaire, mainly the reply to question ``I exercise regularly'', to which a binary yes/no reply is allowed. As a self-declared assessment of action, the variable suffers from the biases endemic to surveys, including acquiescence bias (tendency to reply positively), social desirability bias (tendency to reply in line with perceived expectations), and faulty recall. Participants may also have a unique understanding of the frequency of exercise which may be considered ``regular''. Since we are interested in comparing participants within the study, we make an implicit assumption that the biases and individual noise are uniformly distributed through the population (more on this limitation in the Discussion section). In our data, 8,390 (55.8\%) of respondents indicated they exercised regularly, the rest -- otherwise.

\begin{figure*}[t!]
\raisebox{0.16\height}{\includegraphics[width=0.250\linewidth]{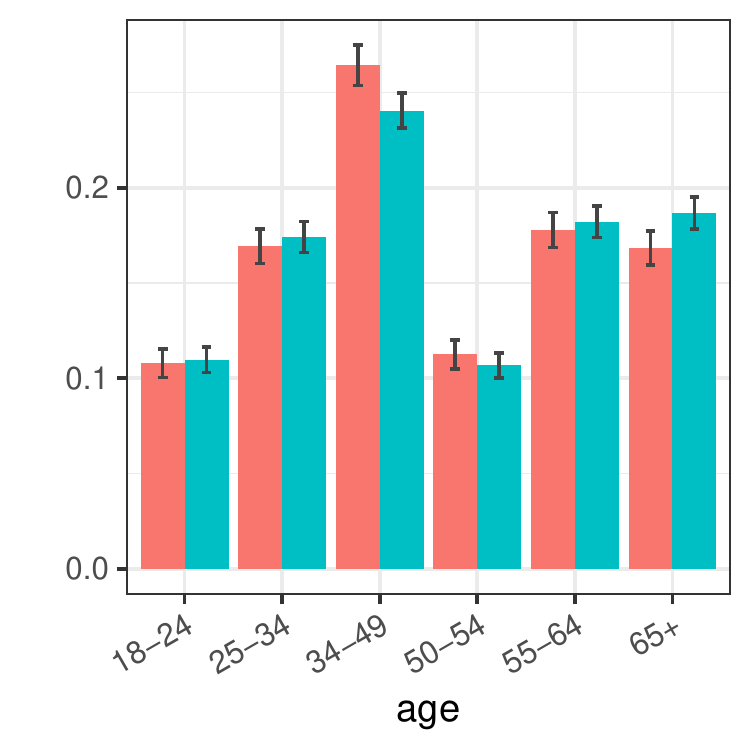}}\hfill
\raisebox{0.15\height}{\includegraphics[width=0.124\linewidth]{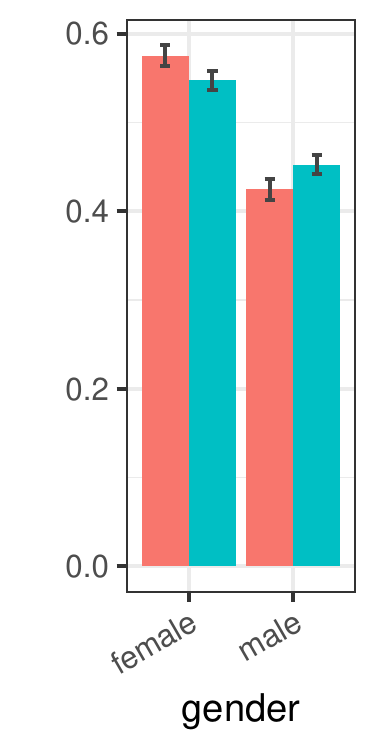}}\hfill
\raisebox{0.00\height}{\includegraphics[width=0.257\linewidth]{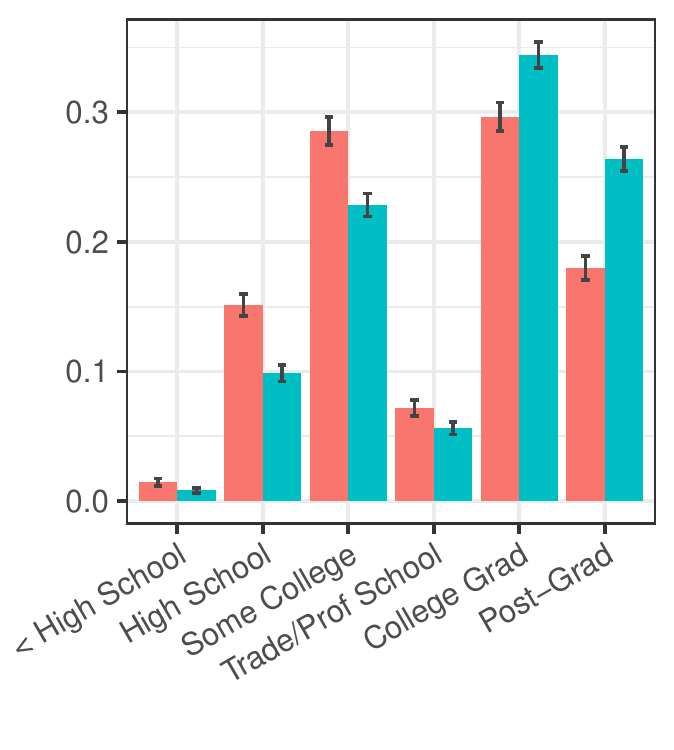}}\hfill
\raisebox{0.08\height}{\includegraphics[width=0.356\linewidth]{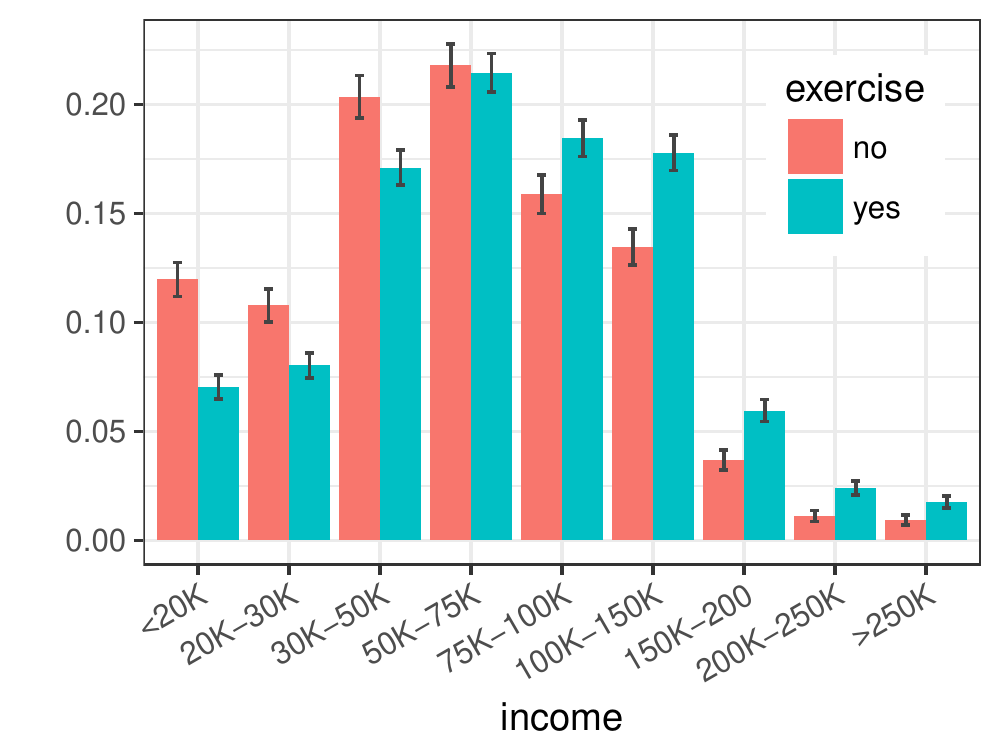}}\hfill
\vspace{-0.5cm}
\caption{Demographic distribution of respondents, broken down by whether they indicated they exercise, with 95\% conf. intervals.
\label{figure:demogplots}}
\end{figure*}

We begin by examining the basic demographic characteristics of the two groups, shown in Figures \ref{figure:demogplots}. As the plots show the 95\% confidence intervals, we can discern some statistically significant differences in the two groups. Mostly, we see no significant age differences, except for in 34-49 range, when it is slightly more likely that the users do not exercise. Similarly, there are slightly more males indicating that they exercise than females. Education and income prove to be a more discerning feature, with college graduates and post-graduates exercising markedly more, and high-school graduates less.
Similarly, the higher income individuals (household income of \$70k or more) reported exercising markedly more than those in the lower brackets (\$50k or less). A similar observation can be made for the wealth variable, with those having a net worth of less than \$50k reporting to be exercising markedly less than those having over \$250k (plot omitted for brevity). These findings echo earlier observed tendency of those in higher income stratum to engage in higher levels of physical activity (as has been described in a literature review \cite{kaczynski2007environmental} and later measured using accelerometers \cite{shuval2017income}).


\subsection{Predicting Exercise}

\begin{table*}[t]
\begin{center}
\caption{Performance of random forest models predicting whether participant indicated exercise, measured using weighted AUROC, along with p-value of significance in difference with Experiment 1 (Basic demographics).}
\label{tbl:exerciseDESCR}
\begin{tabular}{clcc}\toprule
 && Weighted & Significance \\
& Features included & AUROC &  p-value\\
\midrule
    1. & Basic demographics (only gender and age)  & .513 & -\\
	2. & Advanced demographics \Tf  & .616 &  $<$0.0001 \\
	3. & Advanced demographics + values/morals & .623 & $<$0.0001\\
	4. & Advanced demographics + values/morals + value health & .650 & 0.0001 \\
	5. & Advanced demographics + values/morals + value health + domains cat-s  & .654 & $<$0.0001\\
	6. & Advanced demographics + values/morals + value health + app cat-s& .671 & $<$0.0001 \\
	7. & Advanced demographics + values/morals + value health + domains cat-s + app cat-s& .672 & $<$0.0001 \\
	8. & Advanced demographics + values/morals + value health + domains cat-s + app cat-s + H\&F Time & .673 &$<$0.0001  \\ 
	9. & Domains cat-s                         \Tf   & .546 & 0.07 \\
	10. & Advanced demographics + domains cat-s       & .646 & 0.0002\\
	11.  & App cat-s                            \Tf    & .608 & 0.001 \\
	12. & Advanced demographics + app cat-s           & .646 & $<$0.0001\\
	13. & Values/morals                         \Tf   & .575 & 0.002\\
	14. & Values/morals + value health                & .618 & 0.0001\\

\bottomrule
\end{tabular}
\end{center}
\end{table*}

Next, we would like to determine whether it is possible to use this data to predict engagement in physical activity. We formulate this study as a supervised classification problem, aiming at predicting whether participants exercise.
We assess the predictive power of: 

\begin{itemize}[topsep=3pt,itemsep=0pt]
\item Demographics (ethnicity removed due to sparsity)
\item Psychometry: Moral Foundations and Schwartz Basic Human Values, and Emotional Contagion
\item Health-related variables: including replies to survey questions explicitly about health
\item Web domain categories: for both desktop and mobile users
\item Application categories: for mobile users only
\item Rate of usage of Health \& Fitness applications
\end{itemize}

Note that we chose to single out the demographic and survey variables having to do with health and health-related attitudes into their category, as they tend to be highly correlated with exercise. 

Focusing only on our ``Mobile'' dataset for which we have all the above information (n=2620), we train a Random Forest (RF) classifier \cite{Breiman2001} inferring each time from a richer set of predictors as presented in Table~\ref{tbl:exerciseDESCR}. 
The choice of the classifier is motivated by its ability to deal with the sparse web browsing activity data in our dataset, and its performance in previous studies.
We perform ten-fold cross-validation procedure and report the average Area Under the Receiver Operating Characteristic Curve (AUROC) weighted statistic over all folds. 
In the last column of the table, we report the statistical significance obtained by comparing the performance of each model to the basic demographics baseline. 
Note that the random baseline would achieve a weighted AUROC of .50 for all experiments.
For all experiments, our data fusion policy consisted of ``early'' fusion at a feature level, concatenating the different feature vectors for each respondent.

We begin by attempting to predict exercise inferring only on the most basic demographic attributes -- gender and age (Ex.~1 in Table~\ref{tbl:exerciseDESCR}), finding performance to be not much beyond the random baseline. However, adding richer demographic attributes, such as wealth, income, and educational level, significantly improved our prediction to 0.616 (at $p<0.001$).  

Enhancing the baseline model with information about the moral values (Ex.~3), we note an improvement in the performance. The increase is even more pronounced - as expected - with the inclusion of the health-related variables (Ex.~4). 
Adding web browsing domain categories (Ex.~5) slightly improved the model, but it is when including the categories of the applications used (Ex.~6) that the accuracy is increased in a notable way to 0.671.
Including both sets of features - apps and domains categories - (Ex.~7), as well as the average time people used a Health \& Fitness application (Ex.~8) performs statistically identical to Ex.~6, indicating that the mere knowledge of application being opened once is enough. 

Examining the predictive power of each variable type, we find internet browsing domain categories to be the least useful (Ex.~9,10), followed by application usage (Ex.~11,12), and the most valuable (although also most difficult to obtain) the moral values, including those about health (Ex.~13,14) and advanced demographics (Ex.~2) including wealth and income.


Thus, we illustrate the utility of value beliefs in modeling exercise, which combined with demographics and technology usage substantially outperform the baseline demographics model.


\begin{table*}[t]
\begin{center}
\caption{Logistic regression models predicting exercise using demographic (D), values (V), health (H), url domain (U), and app (A) features, applied to users who shared PC activity (P), mobile activity (M) or neither (N). Only features significant at $p<0.01$ level shown (before Bonferroni adjustment), alongside their coefficient estimate and their corresponding p-values (now Bonferroni-adjusted). Confidence levels: $p<0.001$ ***, $p<0.01$ **, $p<0.05$ *.}
\label{tbl:exerciseregression}
\footnotesize
\vspace{-0.2cm}
\begin{tabular}{lrlrlrlrlrlrlrl}\toprule
 & \multicolumn{2}{c}{D (N+P+M)} & \multicolumn{2}{c}{D+V (N+P+M)} & \multicolumn{2}{c}{D+V+H (N+P+M)} & \multicolumn{2}{c}{D+V+H+U (P)} & \multicolumn{2}{c}{D+V+H+U (M)} & \multicolumn{2}{c}{D+V+H+A (M)} & \multicolumn{2}{c}{D+V+H+U+A (M)}  \\\midrule
& \multicolumn{2}{c}{n=15021} &  \multicolumn{2}{c}{n=15021} & \multicolumn{2}{c}{n=15021} & \multicolumn{2}{c}{n=4995} & \multicolumn{2}{c}{n=2260} & \multicolumn{2}{c}{n=2260} & \multicolumn{2}{c}{n=2260}  \\
& \multicolumn{2}{c}{R$_{MF}^2$=0.031} & \multicolumn{2}{c}{R$_{MF}^2$=0.047} & \multicolumn{2}{c}{R$_{MF}^2$=0.074} & \multicolumn{2}{c}{R$_{MF}^2$=0.704} & \multicolumn{2}{c}{R$_{MF}^2$=0.870} & \multicolumn{2}{c}{R$_{MF}^2$=0.868} & \multicolumn{2}{c}{R$_{MF}^2$=0.875}  \\\Tf
(Intercept) & -0.6837 & *** & -1.4740 & *** & 0.1931 &  & -0.0914 &  & 1.1420 &  & 0.6129 &  & 1.0410 &  \\
education & 0.1334 & *** & 0.1372 & *** & 0.1164 & *** & 0.1313 & *** & 0.1243 &  & 0.1294 &  & 0.1239 &  \\
gender & -0.0161 &  & -0.0058 &  & -0.0819 &  & -0.0898 &  & 0.1450 &  & 0.0729 &  & 0.0981 &  \\
income & 0.0561 & *** & 0.0626 & *** & 0.0442 & * & 0.0761 &  & 0.0833 &  & 0.0729 &  & 0.0696 &  \\
parent & -0.1035 &  & -0.1112 &  & -0.1091 &  & -0.0265 &  & -0.3953 &  & -0.3543 &  & -0.4137 &  \\
wealth & 0.1745 & *** & 0.1709 & *** & 0.1518 & *** & 0.1387 & *** & 0.1714 & ** & 0.1453 & * & 0.1743 & ** \\
age & -0.0795 & *** & -0.0858 & *** & -0.0708 & *** & -0.0630 &  & -0.0441 &  & -0.0052 &  & -0.0126 &  \\
marital\_status\_married & -0.0766 &  & -0.0712 &  & -0.0793 &  & -0.3038 &  & -0.1720 &  & -0.2266 &  & -0.1229 &  \\
political\_party\_vote & 0.0253 &  & 0.0385 &  & 0.0304 &  & 0.0396 &  & 0.0403 &  & 0.0465 &  & 0.0507 &  \\\Tf
EC\_Happiness &  &  & 0.0541 & *** & 0.0162 &  & 0.0063 &  & 0.0420 &  & 0.0479 &  & 0.0331 &  \\
EC\_Sadness &  &  & -0.0269 &  & -0.0207 &  & 0.0086 &  & -0.0698 &  & -0.0665 &  & -0.0689 &  \\
MFT\_authority &  &  & -0.0089 &  & -0.0169 &  & -0.0015 &  & -0.0071 &  & -0.0084 &  & -0.0105 &  \\
MFT\_loyalty &  &  & 0.0160 &  & 0.0244 & *** & 0.0144 &  & -0.0043 &  & 0.0007 &  & -0.0027 &  \\
SWV\_achievement &  &  & -0.0594 &  & -0.0594 &  & -0.0442 &  & -0.0788 &  & -0.0675 &  & -0.0776 &  \\
SWV\_benevolence &  &  & -0.1262 & * & -0.1444 & ** & -0.0643 &  & -0.1547 &  & -0.1210 &  & -0.1481 &  \\
SWV\_hedonism &  &  & -0.1418 & *** & -0.1304 & *** & -0.1027 &  & -0.1832 &  & -0.1547 &  & -0.1671 &  \\
SWV\_power &  &  & -0.1655 & *** & -0.1083 & ** & -0.1337 &  & -0.1553 &  & -0.1321 &  & -0.1462 &  \\
SWV\_security &  &  & -0.1871 & *** & -0.2111 & *** & -0.1488 &  & -0.1920 &  & -0.1771 &  & -0.1692 &  \\
SWV\_selfDirection &  &  & -0.0633 &  & -0.0709 &  & -0.0841 &  & 0.0099 &  & 0.0416 &  & 0.0230 &  \\
SWV\_stimulation &  &  & 0.1131 & ** & 0.1221 & *** & 0.1247 &  & 0.1581 &  & 0.1726 &  & 0.1698 &  \\
SWV\_tradition &  &  & -0.1554 & *** & -0.1371 & *** & -0.1389 &  & -0.2205 &  & -0.1608 &  & -0.1980 &  \\
hypocrisy\_increasing &  &  & -0.0253 &  & -0.0304 &  & -0.0182 &  & -0.0619 &  & -0.0557 &  & -0.0578 &  \\\Tf
blood\_pressure\_high &  &  &  &  & -0.4244 & *** & -0.5468 & *** & -0.4903 & * & -0.5326 & ** & -0.5493 & * \\
chronic\_disease &  &  &  &  & -0.4653 & *** & -0.4767 & *** & -0.3628 &  & -0.3190 &  & -0.3262 &  \\
smoker &  &  &  &  & -0.2274 & ** & -0.3870 &  & -0.2631 &  & -0.1437 &  & -0.1423 &  \\
HQ\_1\_health\_plans &  &  &  &  & -0.1579 & *** & -0.1065 &  & -0.1080 &  & -0.0718 &  & -0.0847 &  \\
HQ\_4\_habit\_choice &  &  &  &  & -0.2547 & *** & -0.3755 & *** & -0.4382 & *** & -0.4013 & *** & -0.4491 & *** \\
HQ\_5\_health\_is\_gift &  &  &  &  & 0.0605 & ** & 0.0473 &  & 0.0096 &  & 0.0241 &  & 0.0326 &  \\
HQ\_6\_avoid\_test\_results &  &  &  &  & 0.0576 & ** & 0.0801 &  & 0.0084 &  & 0.0419 &  & 0.0138 &  \\\Tf
Education\_Reference &  &  &  &  &  &  & -0.0040 &  & 0.0022 &  &  &  & 0.0029 &  \\
General\_News &  &  &  &  &  &  & 0.0002 &  & -0.0033 &  &  &  & -0.0035 &  \\
Interactive\_Web\_Applications &  &  &  &  &  &  & 0.0477 &  & 0.9753 &  &  &  & 1.1660 &  \\
Internet\_Radio\_TV &  &  &  &  &  &  & 0.0252 &  & 0.1096 &  &  &  & 0.1118 &  \\
Malicious\_Sites &  &  &  &  &  &  & -0.1538 &  & 0.0120 &  &  &  & 0.0103 &  \\
Motor\_Vehicles &  &  &  &  &  &  & -0.0143 &  & -0.0341 &  &  &  & -0.0353 &  \\
Online\_Shopping &  &  &  &  &  &  & -0.0026 &  & -0.0019 &  &  &  & -0.0021 &  \\
Personal\_Network\_Storage &  &  &  &  &  &  & 0.1282 &  & 0.3129 &  &  &  & 0.2914 &  \\
Personal\_Pages &  &  &  &  &  &  & 0.0366 &  & -0.3409 &  &  &  & -0.3620 &  \\
Recreation\_Hobbies &  &  &  &  &  &  & 0.0161 &  & 0.0213 &  &  &  & 0.0229 &  \\
Search\_Engines &  &  &  &  &  &  & 0.0014 &  & -0.0015 &  &  &  & -0.0017 &  \\
Sports &  &  &  &  &  &  & 0.0061 &  & 0.0131 &  &  &  & 0.0119 &  \\\Tf
total\_web\_visits &  &  &  &  &  &  &  &  & 0.0011 &  & -0.0001 &  & 0.0014 &  \\
total\_app\_time &  &  &  &  &  &  &  &  &  &  & 0.0000 &  & 0.0000 &  \\
ENTERTAINMENT &  &  &  &  &  &  &  &  &  &  & 0.0011 &  & 0.0012 &  \\
HEALTH\_AND\_FITNESS &  &  &  &  &  &  &  &  &  &  & 0.0095 & *** & 0.0095 & *** \\
HOUSE\_AND\_HOME &  &  &  &  &  &  &  &  &  &  & -0.0137 &  & -0.0110 &  \\
LIBRARIES\_AND\_DEMO &  &  &  &  &  &  &  &  &  &  & -0.5137 &  & -0.5396 &  \\
LIFESTYLE &  &  &  &  &  &  &  &  &  &  & 0.0051 &  & 0.0047 &  \\
MAPS\_AND\_NAVIGATION &  &  &  &  &  &  &  &  &  &  & 0.0052 &  & 0.0054 &  \\
MEDICAL &  &  &  &  &  &  &  &  &  &  & -0.0168 &  & -0.0162 &  \\
MUSIC\_AND\_AUDIO &  &  &  &  &  &  &  &  &  &  & 0.0037 &  & 0.0038 & \\\bottomrule
\end{tabular}
\end{center}
\vspace{-0.1cm}
\end{table*}

\subsection{Determinants of Exercise}

In the aim of understanding the contribution of individual variables to whether a person exercises, we employ multivariate logistic regression analysis, building models from demographics-based baseline. The resulting models are shown in Table \ref{tbl:exerciseregression}. In consideration of space, only variables which have a coefficient significant at $p<0.01$ are shown; note however that the p-value markers (stars) shown in the table have been Bonferroni-adjusted to ameliorate the multiple comparison problems. Intuitively, the Bonferroni correction ``punishes'' the significance of features in a larger model, allowing fewer otherwise significant tests pass the adjusted $\alpha$ threshold (as is visible in right-most columns of the table). Also at the top of each model, we show the number of users having non-empty fields available for the data ($n$) and McFadden's $R_{MF}^2$, which relates the (maximized) likelihood value from the current fitted model to null model \cite{mcfadden1973conditional}.

In baseline model using only demographics we find a strong positive relationship between exercise and \emph{education}, \emph{income}, and \emph{wealth}, and a negative one with \emph{age} (with age echoing findings of previous studies  \cite{ernsting2017using,carroll2017uses}). However, the explanatory power of this model is low, according to $R_{MF}^2$. Adding the moral values, we find exercise to be highly related with \emph{happiness} and \emph{stimulation}, and negatively related to \emph{hedonism}, \emph{power}, \emph{security}, and \emph{tradition}. Note the difference between these and values associated with downloading a health app from Section \ref{sec:healthappuse}, now with an negative association with \emph{power} (``attainment of a dominant position'') and positive with \emph{happiness} (latter is well documented to accompany exercise \cite{Lathia2017}).

Next, the addition of health-related variables unsurprisingly produces highly significant coefficients. There is a robust negative relationship between exercise and having \emph{high blood pressure}, some other \emph{chronic disease}, and being a \emph{smoker}. Interestingly, we also find strong effects in the belief statements of respondents. Those who exercise are more likely to agree with the statement ``When I think of making plans for the future, my health is something I strongly consider'' (\emph{HQ 1}), ``We all have a choice about how to lead our lives, and healthy habits are just one example of that'' (\emph{HQ 4}), but are more likely to disagree with ``Health is a gift and there is not much I can do about it'' (\emph{HQ 5}) and ``Sometimes I avoid getting my test results if I think it will be bad news'' (\emph{HQ 6}). These findings underscore the importance of the individual's belief in their innate ability to achieve goals or self-efficacy \cite{litman2015mobile}.


Upon adding the internet browsing data (available for both Desktop and Mobile cohorts), we find the most important domains to be in the areas of \emph{Personal Network Storage} (content management), \emph{Malicious Sites} (may include adult content), and \emph{Interactive Web Applications} (document readers, calendar), though none had a particularly significant p-value after Bonferroni correction.
Though in combination with baseline variables, the model achieves $R_{MF}^2$ of 0.704 and 0.870 for Desktop and Mobile cohorts respectively.

Finally, as we consider the use of applications (listed in caps) for the respondents who shared their mobile activity, we find several application classes beneficial to the model, the most significant of these being \emph{Health \& Fitness}. 
Note that we have also included aggregate technology usage statistics, including a total number of web visits in the observed time, the total app time and (not shown due to insufficient significance) the total web browsing time. However, we find these to be not highly related to exercise behavior.

As the models increase in complexity (see the last two columns of Table \ref{tbl:exerciseregression}), the Bonferroni correction becomes more strict, which reveals the variables most important in modeling exercise: a combination of demographics (wealth), attitude (HQ4 ``Health is a choice''), and technology use (Health \& Fitness Apps).

\section{Discussion \& Conclusions} 


This study is a contribution to an exciting area of research in User Modeling and Personalization into the user-centered design of persuasive technologies and behavior change interventions to improve health and well-being. In the past, psychological features have been used to model healthy shopping habits \cite{Adaji:2018:EGA:3209219.3209253}, cooking \cite{Rokicki:2016:PPG:2930238.2930248}, and engagement in physical activity \cite{Oyibo:2016:DCP:2930238.2930372}. Thus far, existing UMAP literature focused on \emph{personality} traits which could be leveraged for personalization \cite{Adaji:2018:EGA:3209219.3209253,Ferwerda:2017:PTM:3079628.3079693,Chen:2016:IPM:2930238.2930240}. Here, we show that \emph{values} held by the individuals also affect their health behavior. In a sense, it can be seen as a response to a recent study on strategies to encourage diet and physical exercise by Radha et al. \cite{Radha:2016:LRH:2930238.2930251}, who called for the study of factors that could ``explain attitude towards the feasibility level of a recommendation''. 
Here, we illustrate the use of validated, finer-grained value theories in the modeling of technology users, while also contributing an observational technology use and rich demographics.

In particular, insights obtained in this study lead us to recommend the following considerations when designing persuasive technologies for health behavior modification:

\begin{itemize}[topsep=3pt,itemsep=3pt,leftmargin=10pt]

\item Focus on a \textbf{particular activity} to track. We show that applications having most users exercising are marketed for tracking a particular activity, such as walking, running, or cycling. Those centered around particular wearables, for instance, fare less well.

\item Create interventions with \textbf{socioeconomic status} of the users in mind. We find that wealth is one of the greatest determinants of regular exercise. More should be done to understand the barriers of the less wealthy to leading a healthy lifestyle. For instance, it is curious that it is wealth, not income, that remains most predictive in the regression model.

\item Incorporate discussion of \textbf{values} in the interaction or interface. One of the strongest predictors for exercise is the belief that healthy habits are a conscious choice. Making this choice explicit may reinforce this value and encourage engagement.

\item Encourage the expression of \textbf{happiness} and offer emotional rewards. We find people engaging in exercise identifying more strongly with the value of happiness, and although the direction of causation may point both ways, associating positive emotions with physical activity may reinforce the connection. Note that our findings that the value of \emph{power} is negatively associated with exercise suggest that competitions and leader boards may not be appropriate for many users.

\item Use application usage in \textbf{predictive analytics}, in the absence of detailed demographic or value information. In our classification experiments, we find mobile application usage to be much more useful in predicting exercise than internet browsing.

\end{itemize}



As mentioned earlier, the greatest limitation of this study is the reliance on self-reporting when measuring exercise, as many biases are possible. In future studies, we encourage researchers also to obtain permission to gather user physical activity (which can be done unobtrusively via pedometers and heart rate monitors). Further, we realize administering scientifically validated surveys to technology users may be infeasible. However, attempts are being made to detect moral judgments in social media \cite{teernstra2016morality} and associated with images \cite{crone2018socio}, with potential for automatic value detection in the future. Also, although we capture one month of technology use, the study is not longitudinal -- many people may have downloaded and used the apps at some time, but not in our window of observation. Thus conclusions on application adoption/retention should be made while considering the small eventual sample size per application.




Finally, we would like to reiterate the privacy precautions taken in this study, with respondent anonymization, data aggregation, and URL cleaning, which is performed to limit the exposure of participants. Similar precautions should be taken if or when an inference of values or other personal information is performed, such that the user is given greatest possible control over his or her information, as enforced by, for instance, EU General Data Protection Regulation (GDPR).

\section{Acknowledgements}

Y.M. and K.K.,  acknowledge support from the ``Lagrange Project'' of the Institute for Scientific Interchange (ISI) funded by the Fondazione Cassa di Risparmio di Torino (CRT).




\balance

\bibliographystyle{ACM-Reference-Format}


\end{document}